# Comparative study of the real-time optical trapping in the Rayleigh regime for continuous and pulsed lasers

Soumendra Nath Bandyopadhyay [1], Tushar Gaur [2], and Debabrata Goswami*[1,2]

*Email: dgoswami@iitk.ac.in

[1] Department of Chemistry, Indian Institute of Technology, Kanpur, U.P. -208016

[2] Centre for Lasers and Photonics, Indian Institute of Technology, Kanpur, U.P. -208016

# Abstract

Simulating real-time scenarios to predict optical trapping behavior for continuous wave (CW) as well as femtosecond pulsed lasers is a challenging problem. This is especially so, because, for a tightly focused laser, one must also include optical Kerr effect as well as thermal nonlinearity. We show here the distinct differences between CW and femtosecond pulsed laser mediated optical trapping via their effect in scattering and gradient forces in the Rayleigh regime. In our newly developed model, it will also be easier to predict the stability of the trap in real optical trapping scenario as well as provide information regarding choice of solvents, probes, and optical parameters like laser type, power range, wavelength, etc. as may the cases be.

# Introduction

It has been decades since the optical trapping technique marked its beginning [1]. Ever since its development, the technique has been widely accepted for below microscopic studies and controlled manipulations in various fields, such as Biology, Physics, Chemistry, etc. [2]. The capability and usefulness of the technique remain unquestionable, yet one of the questions still prevails regarding the localized heating of the medium about the laser focus during optical trapping experiments. The importance and effects of laser-induced thermal heating in optical trapping have been mentioned numerous times by researchers [2,3] but no significant work has been done in measuring the thermal effects on the force and potential of the trapped particle. The study of these thermal effects in optical trapping becomes even more important to embark on the difference between continuous wave (CW) laser and femtosecond pulsed laser optical trapping. Studies show that force due to a femtosecond pulsed laser is much higher compared to the CW case [4].

It is a well-known fact that when a tightly focused laser beam interacts with the medium, heating takes place due to the absorption of the light intensity [5]. This heating is maximum near the focus. As a result, the focus serves as the heating source in the medium. This leads to temperature rise, which in turn makes the medium rarer resulting in a locally reduced refractive index (of the medium) [6]. The trapping force and potential heavily depend on the refractive index of the medium [7]. Thus, this thermal effect has a great impact on the force and potential of the trapped particle. Different methods [6] have been utilized to determine the increase in temperature near the focus. In the year 2003, E.J.G. Peterman and his associates were able to define an entirely new way of determining the temperature rise near the focus in CW laser optical trapping [6]. However, in the case of ultrafast laser pulses, this method of temperature determination is not approachable due to computational limitations. Thus, a new technique development was needed to understand the thermal effects of ultrafast laser pulses the optical trapping. For this, it is necessary to account for all the thermal effects, both in femtosecond pulsed as well as in CW laser cases. As the laser beam needs to be tightly focused for optical trapping, one should include the thermal nonlinearity coefficient in the equation for forces and potential regarding the trapped particle.

# Theory

In optical trapping, there are two types of forces that act on a trapped particle. One is scattering force (which acts in the direction of the beam propagation), and the other is the gradient force (which pulls the particle towards the higher intensity when the refractive index of the particle is greater than the surrounding medium). For femtosecond pulsed laser, another additional force that comes into the act is the temporal force [7], which is often neglected due to a minimum overall contribution. For a stable trap, the gradient force must overcome all the other forces acting on the particle. Based on the size of the particle, there are various methods of calculating these forces. In this paper, we specifically focus on optical trapping in the Rayleigh regime for which the particle size is much smaller than the wavelength of the trapping beam (λ). As a result, the particle can be considered to be a dipole and Rayleigh scattering theory can be safely utilized to calculate the forces acting on the particle analytically. Forces in the axial direction for this regime are given by [8,9]:

$$F_{scat.}(z = 0, r) = \frac{8\pi n_m (ka)^4 a^2}{3c} \left(\frac{m^2 - 1}{m^2 + 2}\right) I_o \frac{1}{1 + 2\tilde{z}^2} \exp\left(\frac{-2\tilde{x}^2}{1 + 4\tilde{z}^2}\right)$$

$$F_{\text{grad.}}(z=0,r) = -\frac{2\pi n_m a^3}{c}\left(\frac{m^2-1}{m^2+2}\right)I_o\frac{1}{1+2\tilde{z}^2}\left(\frac{\frac{8\tilde{z}}{k\omega_o^2}}{1+2\tilde{z}^2}\right)\left(1-\frac{-2\tilde{x}^2}{1+4\tilde{z}^2}\right)$$

Here $n_m$ and $n_p$ are the refractive index of the medium and the particle respectively, m defines the ratio of refractive index of the particle to the refractive index of the medium, a is the particle radius, k is the wave propagation constant, c is the speed of light, z and x are axial and radial distance from actual laser focus respectively, $\tilde{z}$ and $\tilde{x}$ are reduced co-ordinates and are given by $z/k\omega_o^2$ and $x/\omega_o$, respectively, where $\omega_o$ is the beam radius at the focus. Here $I_o$ is the average intensity for CW and peak intensity for pulse laser given by:

$$I_o = \frac{2P}{\pi\omega_o^2} \text{ and } I_o = \frac{2P}{\pi\omega_o^2 f\tau}$$

As a high intensity tightly focused laser beam passes through a medium, the properties of the medium change due to linear and non-linear interaction with the high energy coherent photons. For the case of linear interaction, the polarization is given simply by:

$$P(t) = \varepsilon_o \chi E(t)$$

However, for the case of nonlinear interaction, the polarization includes higher-order harmonics of the incident field and is now given by [10]:

$$P(t) = \varepsilon_o[\chi^1 E(t) + \chi^2 E^2(t) + \chi^3 E^3(t) \ldots.$$

Here $\chi^n$ is the higher-order susceptibility term, where the superscript n represents the order of nonlinearity. E(t) represents the incident Gaussian electric field for our case. For a material that shows inverse symmetry, the second-order nonlinearity is zero, while the third-order nonlinearity is present. Now, polarization due to the contribution from third-order nonlinearity is given by:

$$P^3(t) = \chi^3 E^3(t)$$

Due to this contribution of third-order nonlinearity, the total refractive index (n) now changes to:

$$n = n_o + n_2 I$$

Here $n_o$ is the linear refractive index of the material, and $n_2$ is the nonlinear refractive index coefficient that arises due to nonlinear optics (NLO) contributions.

There are various methods to determine this nonlinear coefficient of the medium, but due to simplicity and ease of computation, we employed the z-scan method [11] to find out the nonlinear coefficient of the material. In a single beam Z-scan technique, the transmittance is measured via detectors in open aperture and closed aperture setup after passing a tightly focused light beam through a moving sample. Both nonlinear absorption and nonlinear refraction in the material are determined using this technique [11]. Nonlinear absorption is determined by open aperture set up, and nonlinear refraction is determined by closed aperture setup. As the high intensity focused laser beam transmits through a medium, there is a nonlinear phase change that occurs in the beam. This peak nonlinear phase change is given by [11]:

$$\Delta\varphi_o = k_o n_2 I L_{eff}$$

Here $L_{eff}$ is the effective thickness of the material sample, $n_2$ is the nonlinear refractive index coefficient, $k_0$ is wave propagation constant, and I is the average and peak intensity for the case of CW and femtosecond pulsed laser respectively. For a thin sample approximation [11,12] sample thickness (L) is kept less than the Rayleigh range ($Z_o = \pi\omega_o^2/\lambda$), i.e., $L \leq Z_o$, the transmittance at the exit of the sample is given by [11,12]:

$$T(z) = \frac{\int_{-inf}^{inf} P_T(\Delta\varphi_o(t)) dt}{S \int_{-inf}^{inf} P_i(t) dt}$$

Here S is linear transmittance of the aperture, and $P_i(t)$ is the instantaneous power at the input. In our case where we are using a Gaussian beam, this transmittance is given by:

$$T(z) = 1 - \frac{4\Delta\varphi_o \frac{z}{Z_o}}{\left[\left(\frac{z}{Z_o}\right)^2 + 9\right]\left[\left(\frac{z}{Z_o}\right)^2 + 1\right]}$$

Data for transmittance is obtained at the output of the sample using closed and open and closed aperture setup. As $\Delta\varphi_o$ is a function of $n_2$; thus, $n_2$ is obtained by fitting the data obtained to the equation above. Now, $n_2$ obtained in this process can turn out to be both positive as well as negative, depending on the type of nonlinearity present in the medium. In general, non-electronic nonlinearities are present due to thermal effects, such as nonlinear heating. These are slow nonlinearity [13]. A positive value of $n_2$ characterizes an electronically induced nonlinearity while a negative value of $n_2$ characterizes a thermally induced nonlinearity. Usually, in a medium thermally induced nonlinearity dominates over the electronically induced nonlinearity but in the femtosecond pulsed case, the two nonlinearities become comparable.

Now on finding out and adding the nonlinearity of medium and particle, the refractive index of both the particle and the medium changes and is dependent on the intensity of the incident beam. Thus, particle and medium refractive indices are now given as:

$$n_p = n_{po} + n_{p2}I(z,x,t)$$

$$n_m = n_{mo} + n_{m2}I(z,x,t)$$

Here $n_{po}$ and $n_{mo}$ are the linear refractive indices of particle and the medium, respectively, and $n_{p2}$ and $n_{m2}$ are the nonlinear refractive index coefficients of the particle and the medium, respectively. In our case, $n_{p2}$ is due to the electronically induced Kerr nonlinearity, and hence it is positive while $n_{m2}$ is due to thermally induced nonlinearity, and hence it is negative. As the trapped particle is essentially a solid in a fluid medium, the thermal NLO effect will be more prominent for the medium. Here I(z,x,t) is given by:

$$I(z, x, t) = I_o \frac{1}{1 + 2\tilde{z}^2} \exp\left(\frac{-2\tilde{x}^2}{1 + 4\tilde{z}^2}\right)$$

# Results and Discussions

We have experimentally measured thermal nonlinearity constant for both CW and femtosecond pulsed laser cases using Z-scan for water as a medium. The set-up and detailed technique can be found elsewhere [14]. The fitted curve obtained for computing the value of $n_2$ is given in figure 1.

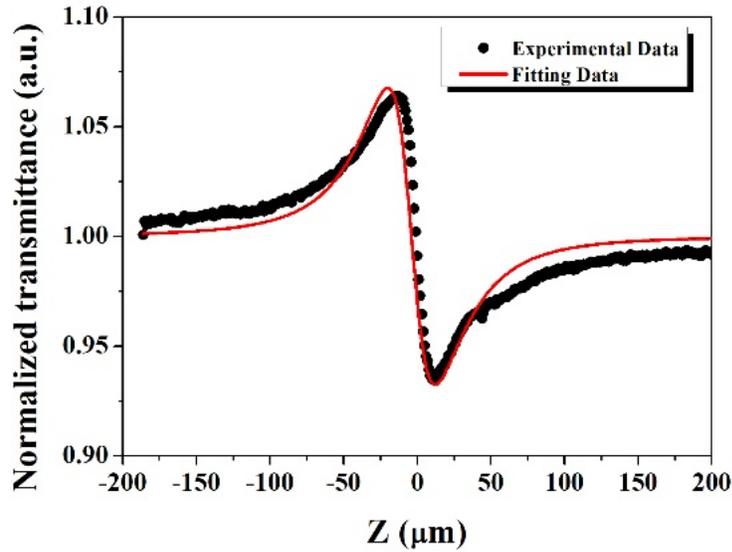

Figure 1. Normalized transmittance curve for experimental (blue) and fitted data (orange) along the y-axis and distance along the z-axis for water.

From here, we obtained $n_{m2} = 1.022\times10^{-18}$ for the case of femtosecond pulsed laser and $1.677\times10^{-13}$ in the case of CW laser. From the data here, we can see that the value of the thermal nonlinearity coefficient in the case of CW laser is much larger than that in femtosecond pulsed laser.

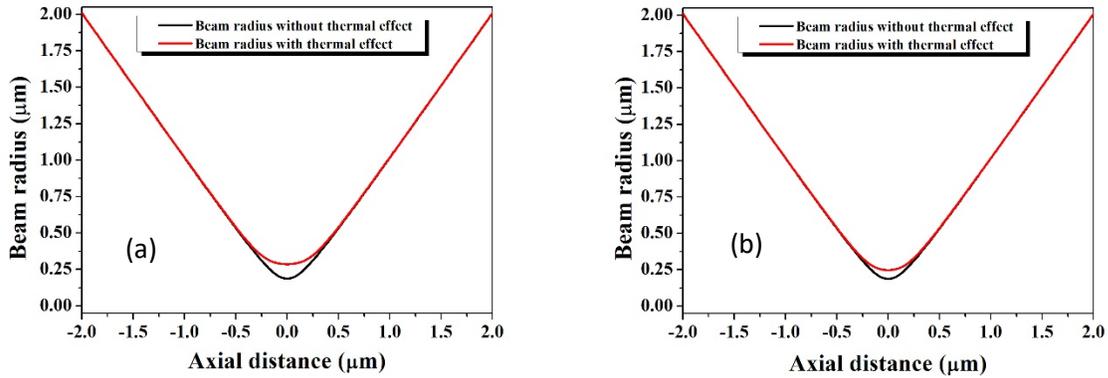

Figure 2. Beam waist for a) CW laser and b) for femtosecond pulsed laser

Thus, from here, we can infer that for the case of CW laser, the thermal effects dominate, which is already known from previous works that temperature rise in the case of CW is higher than that of femtosecond pulsed laser [12,13].

We have taken particles of various sizes ranging from 5nm to 80nm at an interval of 5nm and various average powers ranging from 5 mW to 100 mW both in the case of CW and femtosecond pulsed laser. We have taken the linear refractive indices of polystyrene particle and water medium to be 1.578 and 1.329, respectively [15], and nonlinear refractive index coefficient of the particle is taken to be $5.17\times10^{-17}$ [16-18]. For the case of the femtosecond pulsed laser, the repetition rate (f) is chosen to be 76 MHz and pulse width (τ) is taken to be 150 fs. Here we have used lasers of wavelength 780 nm tightly focused with a numerical aperture of 1.

Thermal effects are first visible in the defocusing effect, which is observed when we plot the beam spot radius over the complete range of length. This defocusing happens due to the heating of medium near the focus which leads to a reduction in the refractive index of the medium, and thus the beam spot radius increases. This is reasonably known that for the same average power, the heating happening due to CW laser is higher than the heating occurring due to femtosecond pulsed laser. As a result, the effects of thermal nonlinearity must be higher in the case of CW laser as compared to that in femtosecond pulsed laser which let us predict that the defocusing in the case of CW laser must be higher than that in femtosecond pulsed laser case that is shown in figure 2.

Figure 2 clearly shows that the shift in the beam spot radius in the case of the CW laser is far more than that in the femtosecond pulsed laser. We calculate the change in the case of CW laser to be 99 nm, while for the femtosecond pulsed laser, it is 61 nm (at 50 mW average power). This result gives us the first indication that the effects of adding thermal nonlinearity must be more significant in the case of CW laser as compared to those in the case of the femtosecond pulsed laser. This is further clear from the previous discussion that the change in the refractive index is going to have a large impact on forces acting on a trapped particle. We have observed these effects by plotting the properly programmed force equations in the Rayleigh regime. The results are generated for particle sizes ranging from 5 nm to 80 nm (which is about the maximum range for Rayleigh particle in 780 nm laser) in 5 nm interval for average powers ranging up to 100 mW. The results are simulated both for CW and femtosecond pulsed laser-based trapping. A few significant plots are shown in figure 3 (force) and figure 4 (potential).

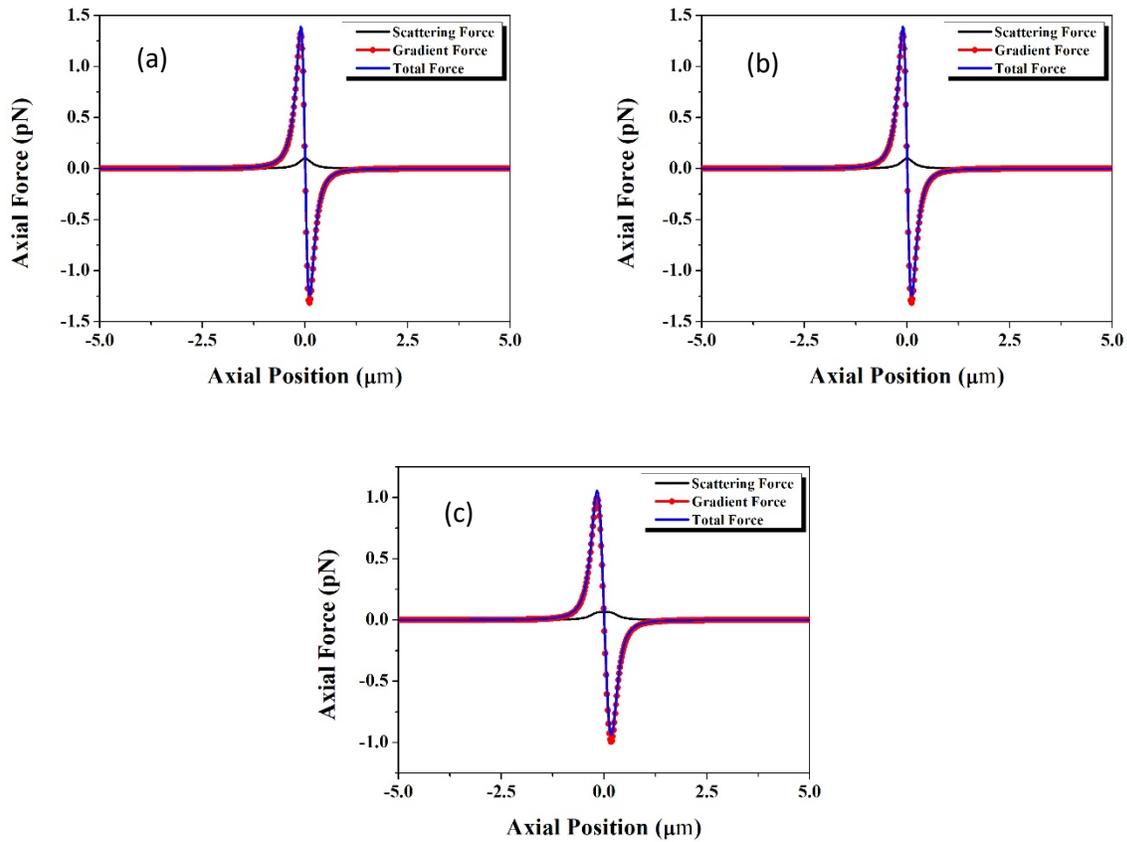

Figure 3: Variation of axial force in different conditions (50 nm radius bead size for 50mW average power CW laser trapping scenario); a) Without the incorporation of any non-linear effects, b) Introducing Kerr nonlinearity, c) inclusion of thermal effects along with Kerr nonlinearity.

Clearly, in the CW laser case, there is no effective change when only the Kerr effect is included. This result is quite an expected one, as in the CW case, the change in refractive index is virtually negligible due to insufficient (average) power. This is where we expect the thermal effect to dominate more significantly over the Kerr one. The effect of thermal nonlinearity is immediately seen in figure 3c as it reduces the magnitude of the total axial force. Also, the change in the scattering force is different as the local refractive index is now changed significantly. For the sake of experimentation, one is more inclined towards knowing the nature of the potentials, which is shown in figure 4.

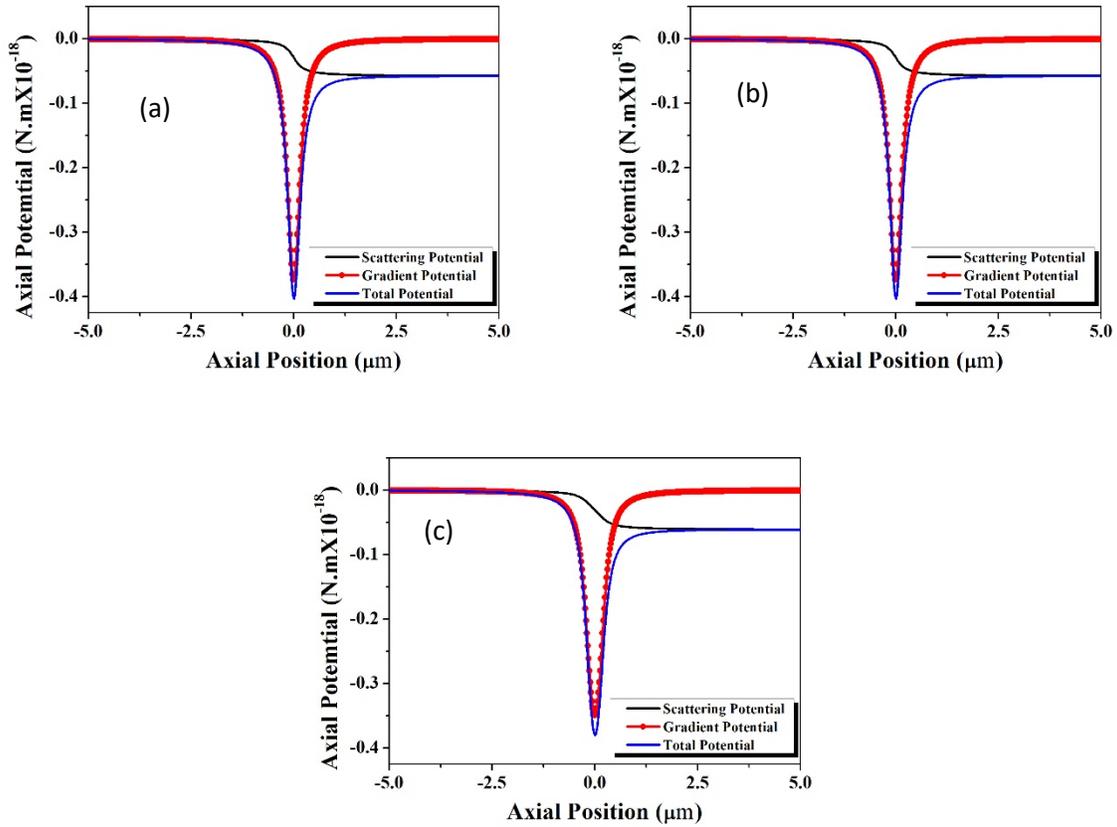

Figure 4: Variation of axial potential in different conditions (50nm radius bead size for 50mW average power CW laser trapping scenario); a) Without the incorporation of any non-linear effects, b) Introducing Kerr nonlinearity, c) inclusion of thermal effects along with Kerr nonlinearity.

In terms of stability of the trapped particle, escape potential is a convenient way to go. We expect that any shallow potential will have lower trapping stability. This result is plotted in figure 5.

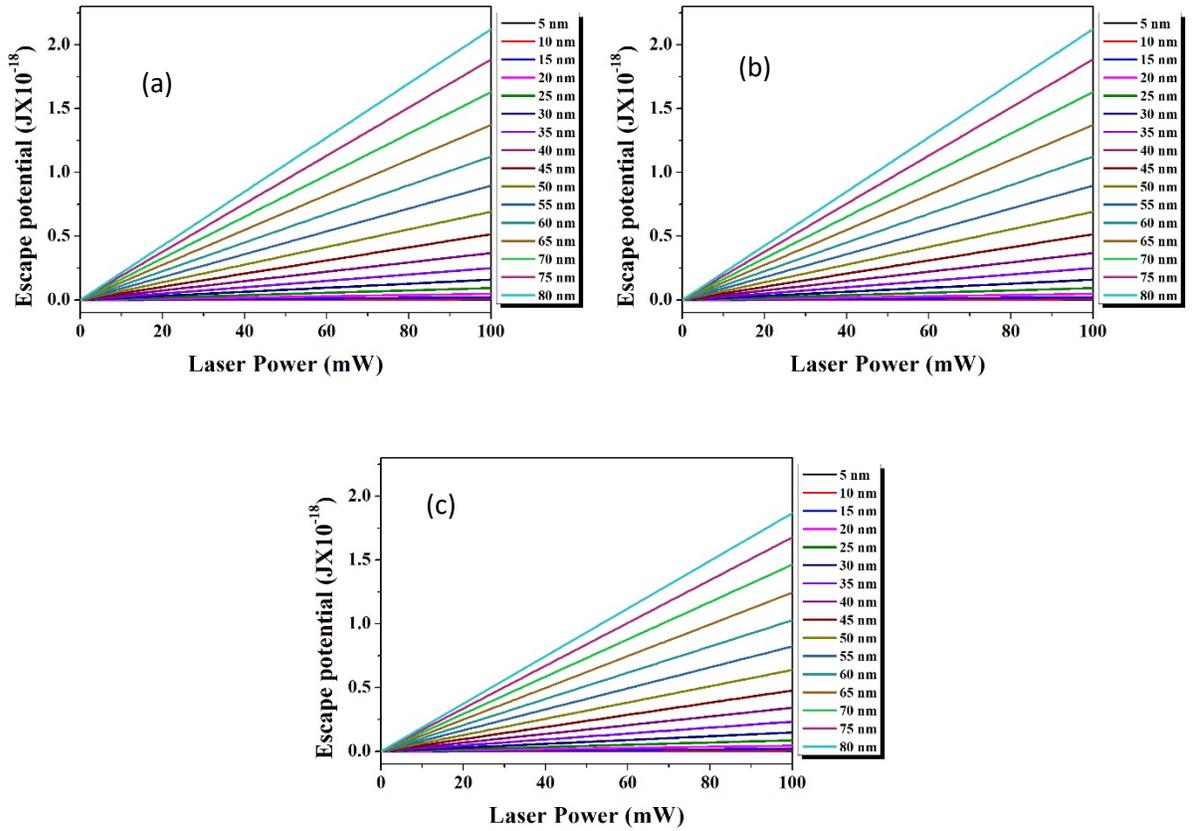

Figure 5: Power dependent variation of axial escape potentials in different conditions in the CW laser trapping scenario; a) Without the incorporation of any non-linear effects, b) Introducing Kerr nonlinearity, c) inclusion of thermal effects along with Kerr nonlinearity.

As expected from the change of optical forces, the potentials differ only when the thermal effects are active. For the Rayleigh region, the particle size is very small, and the thermal property of the solvent (which occupies the major portion of the focal volume) is very important. As the average laser power increases, this effect becomes more and more significant. A more interesting fact is that the size-dependent change is more significant, which indicates that a researcher must choose a power range and size range correctly for experimental purposes. The effect of the radius of the bead is significant because though the beads are very small, but with an increase in radius, the focal volume coverage is increasingly different. Hence the nonlinear effects and overall change in force (or potential) are significantly different. This result is shown in figure 6.

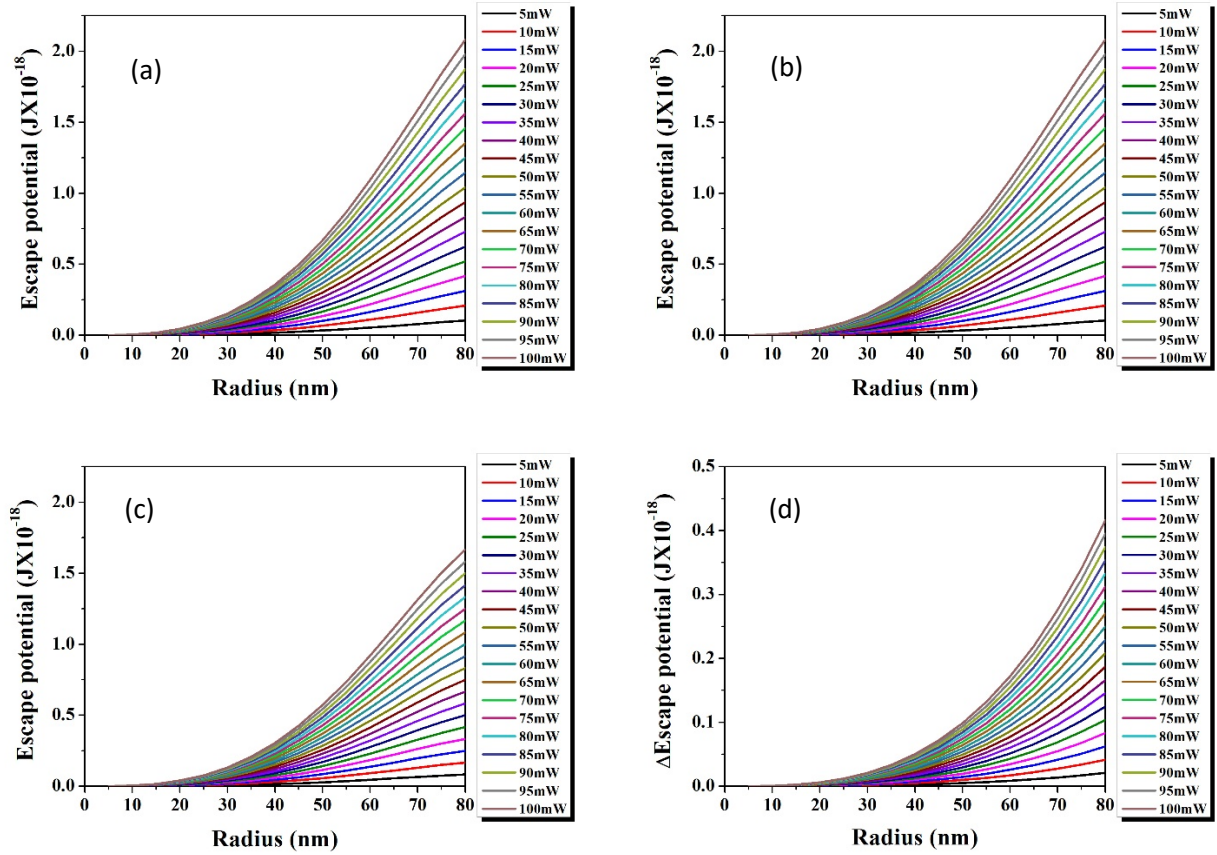

Figure 6: Size-dependent variation of axial escape potentials in different conditions in CW laser trapping scenario; a) without Kerr or thermal effects, b) with Kerr effect, c) Kerr and thermal effect, d) decrease in escape potential due to thermal and Kerr effect relative to the pure one.

When considering the power-dependent variation of the escape potential (of a fixed radius bead), there is a maximum of about ten times greater decrease when the thermal effect is introduced, which is expected as thermal destabilization of the trap should occur. The size dependency is presumably of larger significance, as can be seen from figure 6d. At higher power, more bending is observed.

The case of femtosecond pulsed laser trapping is tricky theoretically, as there can be several nonlinear effects to be considered. Earlier works [19, 20] have considered Kerr effects, but we show here that the thermal effect is even more critical. Figure 7 plots the theoretical axial force generation for the femtosecond pulsed lase mediated trapping phenomenon for different effects being considered.

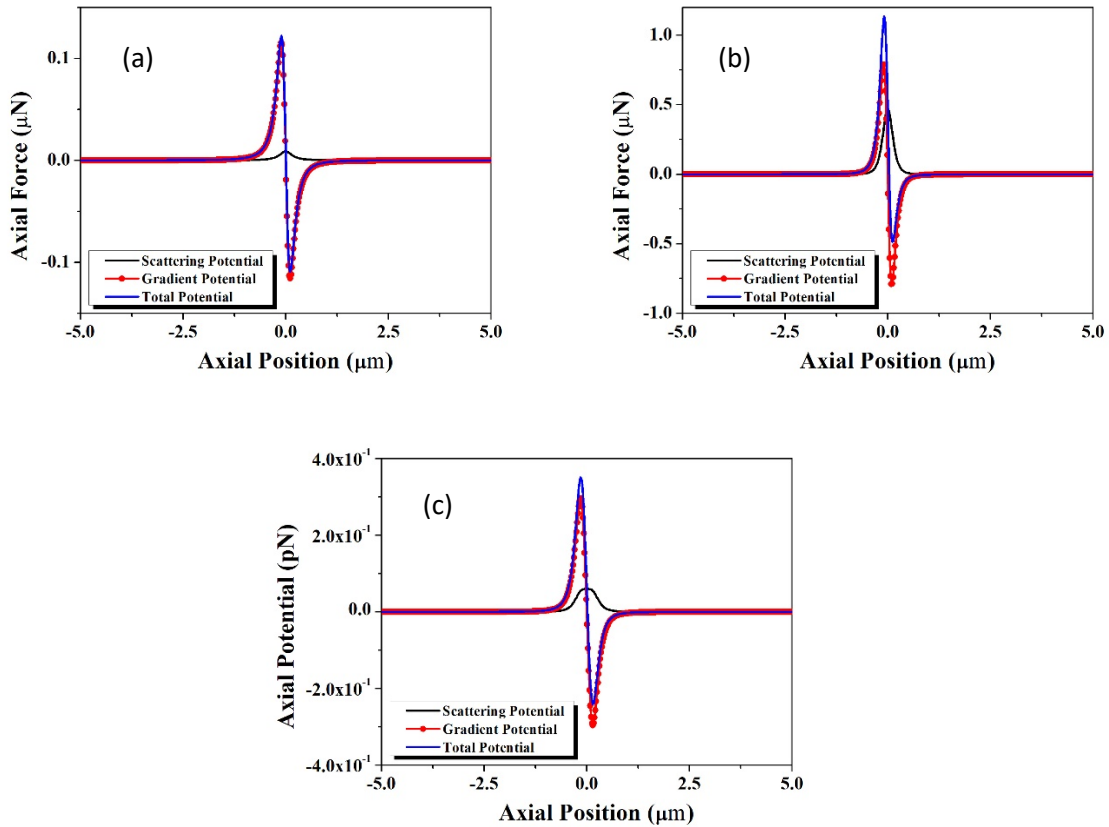

Figure 7: Axial force (50 nm radius bead with 50 mW average power) in different conditions in femtosecond pulsed laser trapping scenario; a) without any nonlinear or thermal effects, b) Introducing Kerr effect, c) incorporation of thermal effects

As seen in figure 7, the Kerr effect increases the total force by almost a factor of 10 (figure 7a and 7b), but the addition of thermal effect reduces the net effect. So, when we approach the real scenario where both Kerr and thermal effects are there, there will an increase in the force, which is the combined effect of Kerr nonlinearity and the thermal effect. The net stability would be guided by the respective potentials, which are plotted in figure 8.

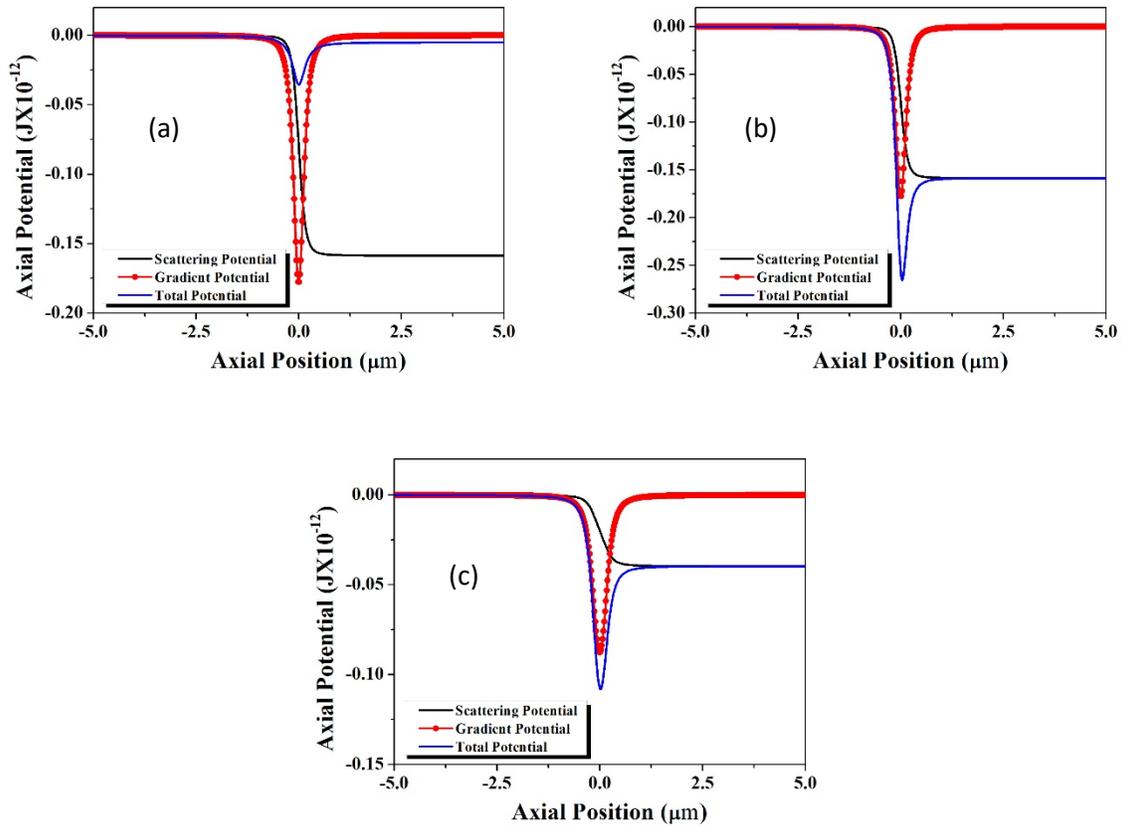

Figure 8: Variation of axial potential in different conditions (50 nm radius bead size for 50 mW average power femtosecond pulsed laser trapping scenario): a) Without the incorporation of any non-linear effects, b) Introducing Kerr nonlinearity, c) Inclusion of thermal effects along with Kerr nonlinearity.

The potential curve in figure 8a shown a very small trapping possibility, whereas the introduction of the Kerr effect shows greater probable trapping under the femtosecond pulsed case. However, when thermal effects are introduced, we get a relatively destabilized trapping potential. The power-dependent escape potentials are plotted in figure 9.

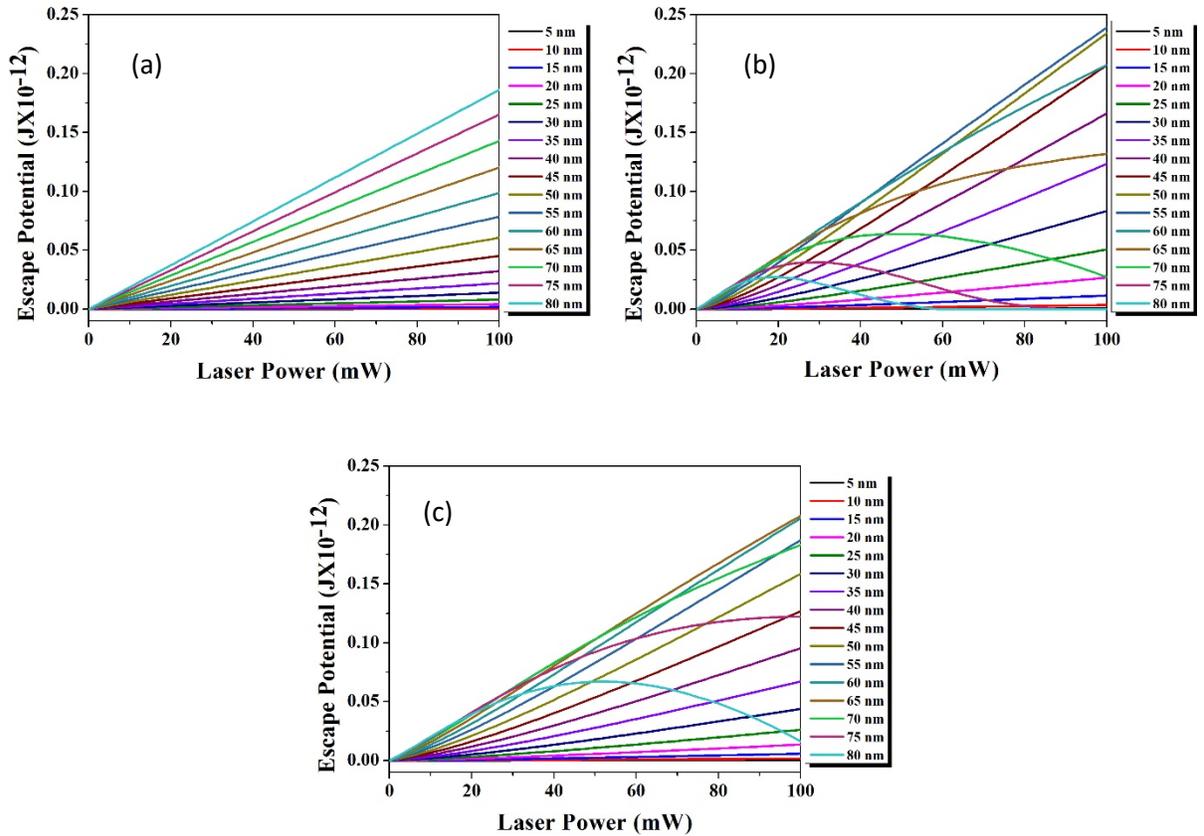

Figure 9: Size-dependent variation of axial escape potentials in different conditions in femtosecond pulsed laser trapping scenario; a) without any effects b) with Kerr effect, c) Introducing thermal effect.

When no effects are used, the escape potential varies linearly with power as expected. When the Kerr effect is used, the curve starts to bend from a 55 nm radius bead onwards. When the thermal effect is introduced, the overall escape potential gets lowered, but now we have a larger range of power available for larger radius beads (bending starts from 65 nm bead and even for 80 nm bead, the escape potential is now not zero at 100 mW).

The relative change of escape potential due to the introduction of the Kerr effect is shown in figure 10a. The rate of change of escape potential is less at higher average laser power. In figure 10b, the relative changes (for escape potential without any effects) are shown when both Kerr and thermal effects are being introduced.

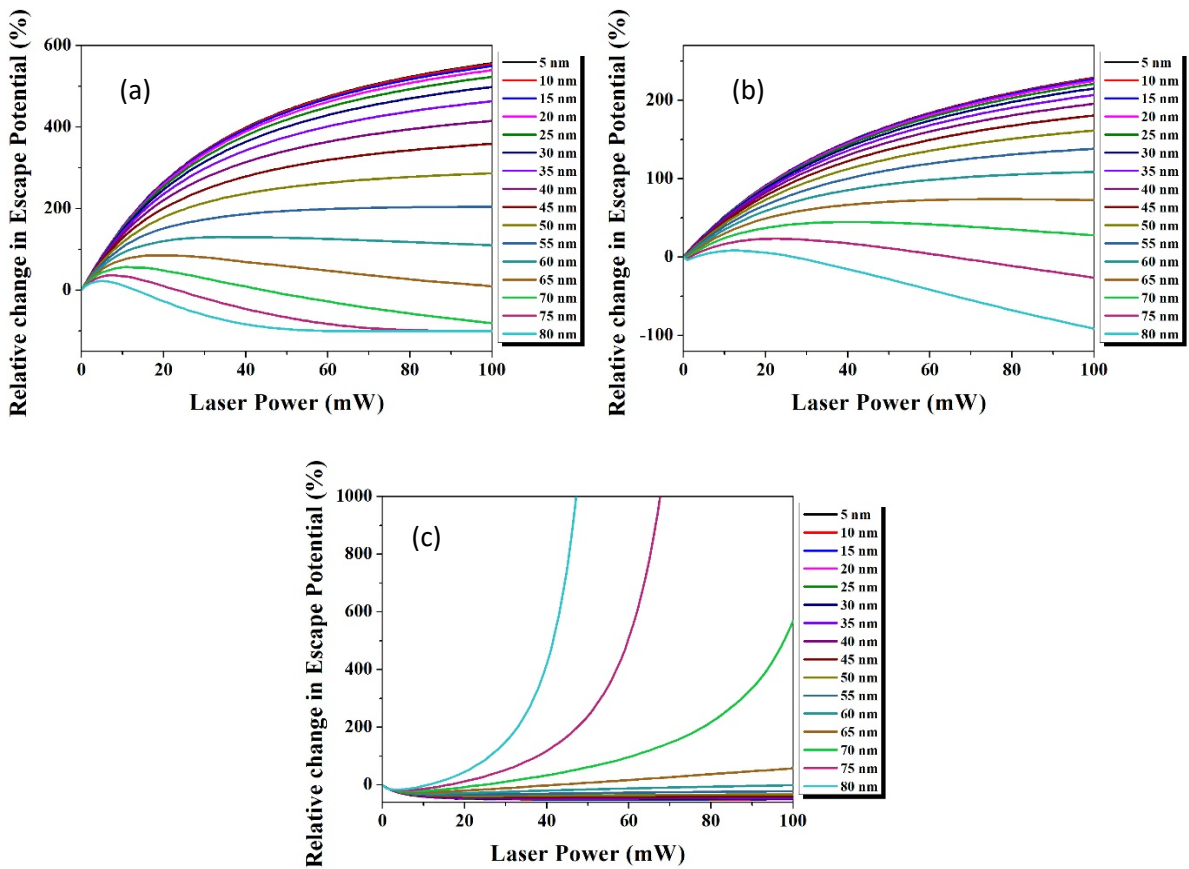

Figure 10: Relative variation of axial escape potentials in different conditions in femtosecond pulsed laser trapping scenario; a) relative increase of escape potential when the Kerr effect is introduced only; b) relative change when both effects are introduced (for pure potential); c) relative change of potential when the thermal effect is introduced for Kerr-only effect.

Here again, at higher laser power, the relative increase is compromised due to the inclusion of thermal effects. In figure 10c, the curves for large radius shoots high as the inclusion of thermal effects stabilize the trap (noticing that change due to Kerr effect makes the escape potential go to almost zero for large radius particles).

Further, we show the radius dependent change in figure 11 as this looks more interesting, and one must choose the right sized bead for experimental purposes.

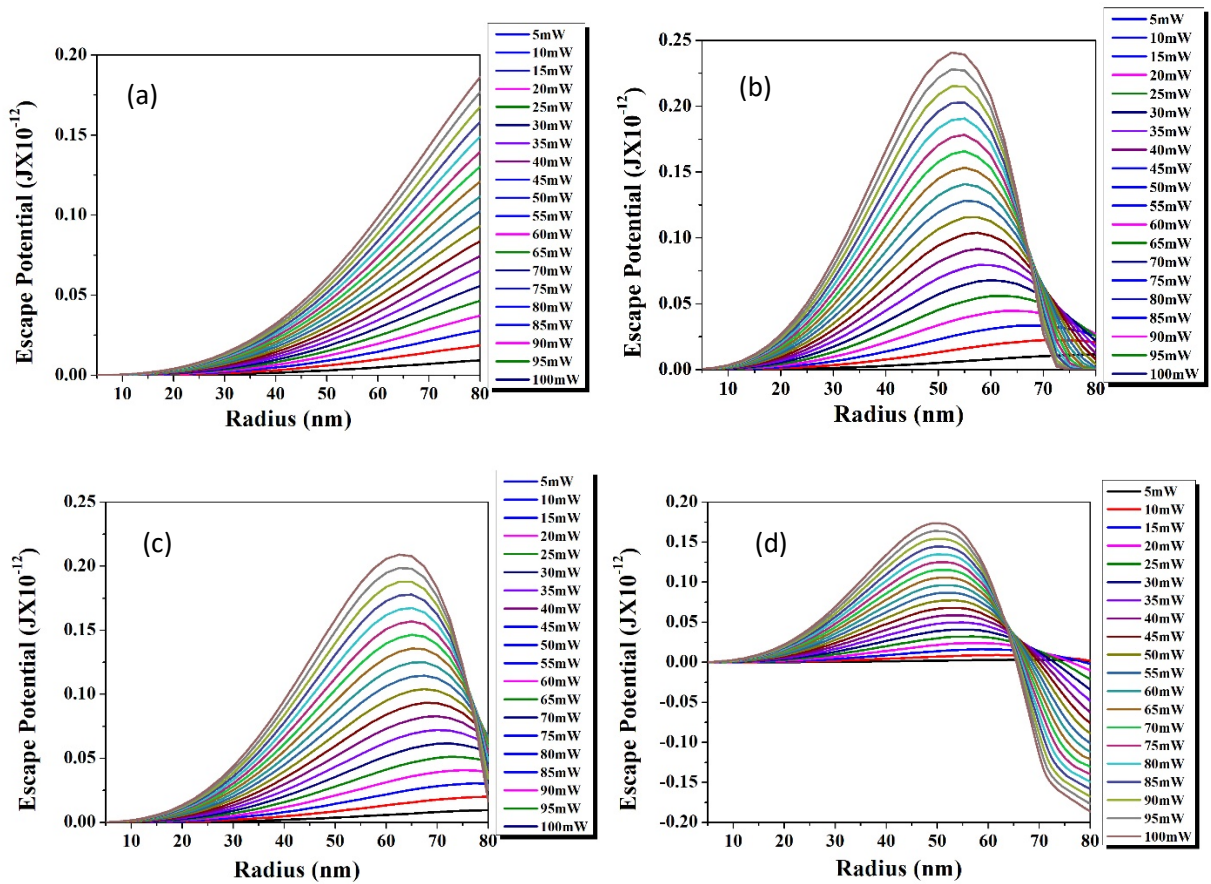

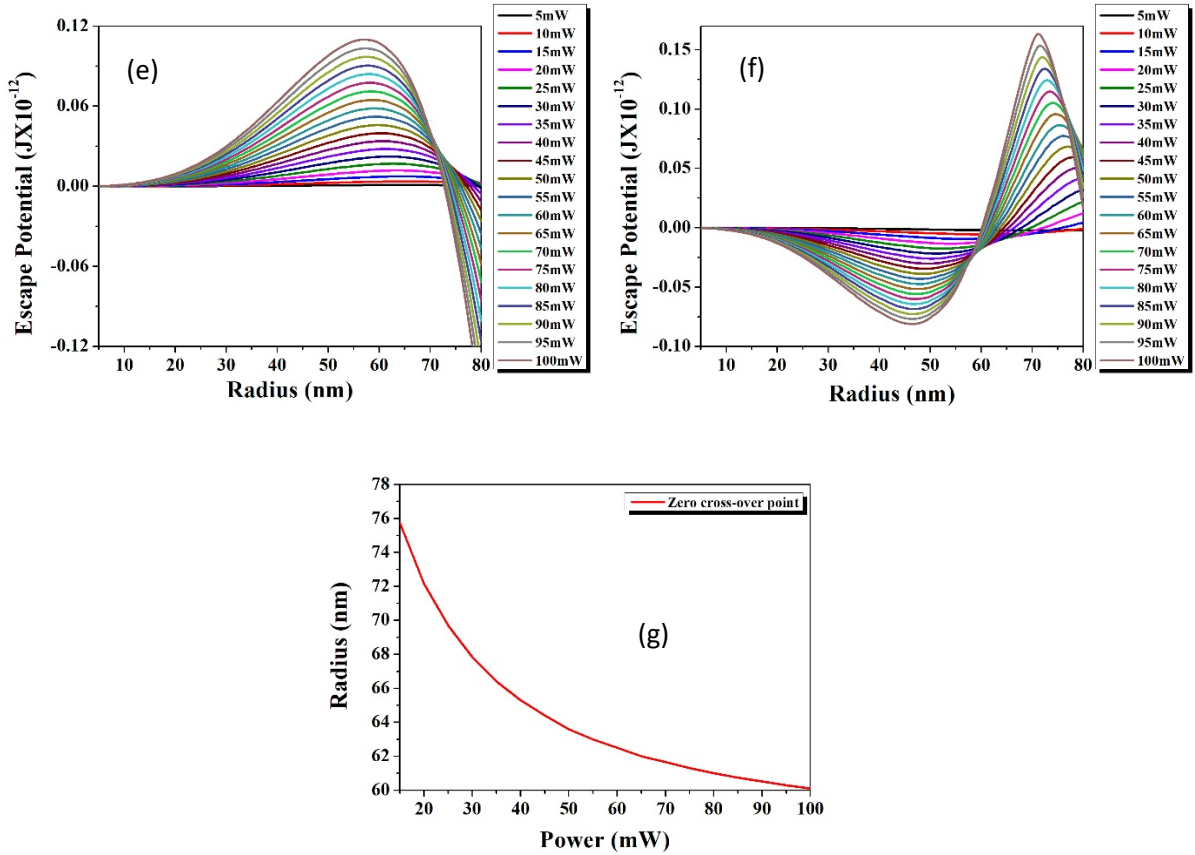

Figure 11: Bead radius dependent variation of axial escape potentials in different conditions in femtosecond pulsed laser trapping scenario; a) increase of escape potential without any effects introduced, b) change in escape potential when only Kerr effect is introduced, c) change in escape potential when thermal effects are introduced along with Kerr effect, d) change of escape potential when Kerr effect is introduced, e) change of escape potential when both effects are introduced over pure escape potentials, f) change of escape potential when thermal is introduced over Kerr effect, g) plot of zero cross-over points of the graph (f).

The last plot of Figure 11g is interesting as it shows the nullifying effect of thermal over Kerr nonlinearity, at these points, the Kerr effect and the thermal effects are exactly balanced out, and one should expect the 'pure' behavior of the bead under femtosecond pulsed trapping scenario.

# Conclusion

Thermal effects have a significant role to play in optical forces. We have introduced linear and nonlinear thermal effects to predict the nature and stability of optical traps both in CW and femtosecond pulsed trapping scenario. The thermal effect for CW mediated optical trapping is quite different from the femtosecond pulsed one. The actual beam spot size plays an active role in this case, which affects the nonlinear refractive index generated near the focal spot. Again, another difference arises due to Kerr nonlinearity, which changes effectively only for the femtosecond pulsed optical trapping as the instantaneous intensity is more important in this case. We see that the size-dependent destabilization is comparatively more, which can be attributed mainly to the thermal effect of laser on the system. In this paper, we have chosen the commonly used polystyrene bead system in water for the broad application of our newly developed theoretical model into experiments. The addition of unavoidable thermal effects will empower us to simulate experimental results in the Rayleigh domain. The power dependence and size dependence on the stability of the optical trap will guide the choice of beads and experimental conditions beforehand. Further, the modeling of Brownian motion for actual experimental conditions along with our newly developed theory, may provide actual prediction and simulation of the actual experimental optical trapping situation.

# Acknowledgment

Soumendra thanks CSIR India for providing support for Ph.D. We thank ISRO, SERB, and MeitY, Govt. of India, for providing funding. For changes in the linguistic style, formatting, as well as for assistance with copy-editing, we thank Mrs. S. Goswami of the Femtolab at IIT Kanpur.